\documentstyle[12pt]{article}
\def\Box{\hbox{$\sqcup$\kern-0.66em\lower0.03ex\hbox{$\sqcap$}}}

\begin{document}
\begin{titlepage}
\begin{flushright}
IFUP--TH 48/98\\
\end{flushright}
\vskip 1truecm
\center{\bf REGGE-LIOUVILLE ACTION FROM GROUP THEORY}\footnote{To
appear in the Proceeding of the 6th International Conference on ``Path
Integral from peV to TeV'' Florence, Italy 26-29 August 1998.}   

\bigskip

\centerline{P. MENOTTI}

\centerline{\it Dipartimento di Fisica della Universit\`a di Pisa} 
\centerline{\it via F. Buonarroti 2, 56100 Pisa, Italy} 
\centerline{\it E-mail: menotti@mailbox.difi.unipi.it}
\vskip 2cm

\begin{abstract}
We work out the constraints imposed by $SL(2C)$
invariance for sphere topology and modular invariance for torus
topology, on the discretized form of Liouville action in Polyakov's
non local covariant form. These are sufficient to completely fix the
discretized action except for the overall normalization
constant and a term which in the continuum limit goes over to a
topological invariant.  
The treatment can be extended to the supersymmetric case.
\end{abstract}
\end{titlepage}

\noindent 
In a seminal paper \cite{ap} Polyakov showed that due to
the functional integration measure induced by the De Witt supermetric,
two dimensional gravity, which is trivial at the classical level, at
the quantum level becomes a highly non trivial theory. The mean value
of a diffeomorphism invariant quantity ${\cal F}$ in two dimensional 
gravity in the conformal gauge is given by \cite{ap,apmm} 
\begin{equation}\label{papmn}
\int {\cal F}\; {\cal D}[\sigma]
\sqrt{\frac{\det'(P^{\dag}P)}{\det(\phi_{a},
    \phi_{b}) \det(\psi_{k},\psi_{l})}}
\det ( \psi_{m},\frac{\partial g}{\partial \tau_{n}} )
\; \frac{d\tau_i}{v(\tau)}.  
\end{equation}
$P^\dagger P$ is the conformal Lichnerowicz-De Rahm operator, $\phi_a$
are the conformal Killing vector fields and $\psi_k$ the Teichm\"uller
deformations. The square root appearing in eq.(\ref{papmn}) gives rise
to the well known Liouville action; the terms after the square root are also
explicitly computable and are absent for sphere topology. ${\cal
D}[\sigma]$ is the functional integration measure induced by the distance
\begin{equation}
(\delta\sigma,\delta\sigma) = \int d^2\omega e^{2\sigma}
\delta\sigma\delta\sigma 
\end{equation}
and is not the translationally invariant measure which usually occurs
in field theory.

The motivation of this investigation is to give the functional
integral (\ref{papmn}) a concrete 
definition, i.e. something that at least in principle could be put on a
computer. The usual procedure in defining a functional integral is to
reduce the number of degrees of freedom to a finite one, then perform
the integration and finally take the limit when the number of degrees
of freedom goes to infinity. One can imagine several ways to break down
the number of degrees of freedom to a finite one, e.g. retaining a
finite number of terms in a mode expansion
but in doing so one easily breaks some 
basic invariances of the theory which are invariance under the
$SL(2C)$ group for sphere and modular invariance for torus
topology; these are remnants of the original diffeomorphism invariance of the
action before the gauge fixing. The Regge scheme which consists to
consider surfaces which are everywhere flat except for a finite number
of points has the advantage to preserve exactly the above mentioned
symmetries. 
 
A complete inclusion of the Liouville term 
which, we recall, arises from the De Witt supermetric, appears necessary
because in two dimensions the integration measure is the only term
which plays a dynamical role; this was also indicated by the failure of naive
Regge integration measure to reproduce the KPZ critical indices
\cite{janke}. For the role of the analogous of the Liouville action in
$D$-dimensional gravity see \cite{mm}.

Finally as the Regge scheme interpolates between the continuum
theory and the more schematic dynamical triangulation models
\cite{discrete} it could 
serve to clarify ``the still mysterious relation between the
continuous 2d quantum gravity and the matrix models \cite{poly}''. 

The exact computation of the Liouville action for Regge surfaces with
the topology of the sphere was
given in \cite{pmppp} and the result is
\begin{equation}\label{reggeliouville}
S_l=  \displaystyle {\frac{26}{12} \left\{ \sum_{i,j\neq i}
    \frac{(1-\alpha_{i})(1-\alpha_{j})}{\alpha_{i}} \log |\omega_{i} -
    \omega_{j}| + \lambda_{0} \sum_{i} (\alpha_{i} - \frac{1}{\alpha_{i}})
    + \sum_{i} f(\alpha_{i}) \right\}}
\end{equation}
$\alpha_i$ are the conical openings of the singularities ($\alpha=1$ is
the plane) and $\omega_i$ are complex number describing the locations
of the singularities on the complex plane closed by the point at
infinity (Riemann sphere) and $\lambda_0$ the scale parameter.  $f$ is
a know function and the last term reduces to a topological invariant
in the continuous limit.
The procedure for proving eq.(\ref{reggeliouville}) employs the heat
kernel technique; the Riemann-Roch 
theorem is used to fix the correct self adjoint extension of the
operators $P^\dagger P$ and $P P^\dagger$ on a conical
singularity. Similar procedure furnishes the discretized Liouville
action for torus topology\cite{pmppp}. 

In \cite{pm} the question was addressed whether a simpler procedure exists
for deriving eq.(\ref{reggeliouville}). More explicitly: which are the
constraints the $SL(2C)$ invariance for the sphere topology, or modular
invariance for the torus topology imposes on the discretized action?

Let us starts from Polyakov's non local covariant form for the Liouville
action on the continuum
\begin{equation}
S_l = \frac{26}{96\pi} \left\{\int  d^{2}\omega \, d^{2}\omega'
\;(\sqrt{g}  R)_{\omega }
\frac{1}{\Box}(\omega, \omega')(\sqrt{g} R)_{\omega'}   
 - 2 (\log {A\over A_0}) 
\int \: d^{2} \omega\,\sqrt{g} R \right\}
\end{equation}
and look for a discretized structure where the curvature is
localized at a finite number of points. Referring to sphere topology
the most general form we have is
\begin{equation}\label{general}
\sum_{i,j} K_{ij}[\alpha] \log|\omega_i-\omega_j| + B(\lambda_0,\alpha).
\end{equation}
Imposition of invariance under dilatations restricts the structure of
the bulk term to $B = -\lambda_0 \sum_{i,j}K_{ij}[\alpha] +F[\alpha]$.
Further imposition of invariance under the general $SL(2C)$
transformation gives rise to the following equation
\begin{equation}
2\sum_{j} h(\alpha_i,\alpha_j) (1-\alpha_j)=
\sum_{m,n}(1-\alpha_m)(1-\alpha_n)h(\alpha_m,\alpha_n)
\end{equation}
where we have defined $K_{ij}[\alpha] =
(1-\alpha_i)(1-\alpha_j)h(\alpha_i,\alpha_j)$. Such equation has 
to be solved under the Gauss-Bonnet constraint $\sum_i(1-\alpha_i)
=2$. It is possibile to show \cite{pm} that the unique solution is
\begin{equation}
h(\alpha_1, \alpha_2) = \left(\frac{1}{\alpha_1}+\frac{1}{\alpha_2}\right)
\frac{h(1,1)}{2} 
\end{equation}
which substituted into eq.(\ref{general}) gives rise to the action
eq.(\ref{reggeliouville}) except for the the overall normalization
constant and the function $F[\alpha]$. We know from the
structure of the heat kernel 
technique, without actually performing the derivation, that
$F[\alpha]$ has the local structure $F[\alpha]= \sum_i f(\alpha_i)$ which
in the continuum limit 
reduces to a topological invariant. The normalization constant 
cannot be forecasted by group theory alone. In
fact the same group theoretical argument can be applied to derive also
the
usual trace anomaly for a scalar field. The value $26/12$ can be
borrowed from perturbation theory; it is worth 
recalling however that such value is non perturbative, and it is an
outcome of the non perturbative the heat kernel derivation\cite{pmppp}.

One can also derive the discretized Liouville action for torus
topology exploiting invariance under the modular group; as the modular
group is a discrete group and thus in a sense weaker than a Lie group,
two further assumptions are needed. These
are 1) the independence of the divergent behavior when two
singularities collide 
of the topology of the surface; 2) a regularity property of the
dependence of the action on the modulus. The result agrees with the
complete derivation \cite{pmppp}.

It is interesting that this group theoretical approach can be extended
to the supersymmetric case. The treatment was done in collaboration
with G. Policastro \cite{pmgp}. Here the super Regge surface will be flat
everywhere except for a finite number of points in superspace
\cite{foerster}. For sphere topology the superconformal factor takes
the form
\begin{equation}
\Sigma ({\bf z}) = \frac{1}{2} \left (\sum_i (\alpha_i-1)\ln [(z-z_i
+\theta\theta_i)(\bar z-\bar
z_i-\bar\theta\bar\theta_i)]+\lambda_0\right ).
\end{equation} 
The procedure works very similarly as in the ordinary case with
the result given by eq.(\ref{reggeliouville}) with
$|\omega_i-\omega_j|$ replaced by ${\bf z}_{ij} \bar{\bf z}_{ij}$
being ${\bf z}_{ij}$ the superdisplacement ${\bf z}_{ij} = z_i - z_j +
\theta_i\theta_j$ and the constant $26/12$ replaced by $10/32$ which
again is borrowed from perturbation theory.
Similar explicit result is obtained for the even spin structures on the
supertorus, while for the odd spin structure on the supertorus the
result is determined up to a second order polynomial in the
coefficients of the zero modes \cite{pmgp}. 
Finally to complete the treatment of the supersymmetric case one
should compute explicitly the functional integration measure for the
superconformal factor.  

We recall that in the ordinary case such a measure is explicitly known
in the form 
of a $3N\times3N$ determinant \cite{pmppp}. Such determinant satisfies exactly
$SL(2C)$ or modular invariance. The analytical handling of such a term
is an important problem and should clarify the relation between
the original Weyl invariant discretized integration measure and the
simpler translational invariant measure.

\bibliographystyle{plain}

\end{document}